\documentclass[12pt]{article}
\usepackage{a4wide}
\usepackage{amssymb}
\begin{document}
{\renewcommand{\thefootnote}{\fnsymbol{footnote}}
\begin{center}
{\LARGE  Canonical derivation of effective potentials}\\
\vspace{1.5em}
Martin Bojowald\footnote{e-mail address: {\tt bojowald@gravity.psu.edu}}
and Suddhasattwa Brahma\footnote{e-mail address: {\tt sxb1012@psu.edu}}
\\
\vspace{0.5em}
Institute for Gravitation and the Cosmos,\\
The Pennsylvania State
University,\\
104 Davey Lab, University Park, PA 16802, USA\\
\vspace{1.5em}
\end{center}
}

\setcounter{footnote}{0}

\begin{abstract}
 A canonical formulation of effective equations describes quantum corrections
 by the back-reaction of moments on the dynamics of expectation values of a
 state. As a first step toward an extension to quantum-field theory, these
 methods are applied here to the derivation of effective potentials around a
 homogeneous vacuum expectation value of scalar fields. A comparison with the
 standard Coleman--Weinberg potential shows that the new methods correctly
 include all relevant quantum corrections. At the same time, the effective
 potential is shown to be correct also for non-Fock and mixed states. Several
 explicit results are derived in models of interacting scalars and fermions.
\end{abstract}

\section{Introduction}

Canonical effective methods \cite{EffAc,Karpacz,HigherTime} have so far
produced several results in quantum-mechanical systems which can serve as
analogs of effective potentials in quantum field theory. In this setting, one
introduces an effective Hamiltonian $\langle\hat{H}\rangle$ by taking the
expectation value in an arbitrary state parameterized by the expectation
values $(\langle\hat{q}\rangle,\langle\hat{p}\rangle)$ and moments
\begin{equation} \label{Moments}
 G^{a,b}= \left\langle (\hat{p}-\langle\hat{p}\rangle)^a
(\hat{q}-\langle\hat{q}\rangle)^b
 \right\rangle_{\rm Weyl}
\end{equation}
of basic operators $\hat{q}$ and $\hat{p}$ --- assuming for now a single
canonical pair. The subscript `Weyl' indicates totally symmetric ordering of
the factors. For the harmonic oscillator, for instance, we have
\begin{equation} \label{EffHam}
 \langle\hat{H}\rangle =\frac{\langle\hat{p}\rangle^2}{2m}+
 \frac{1}{2}m\omega^2 \langle\hat{q}\rangle^2+ \frac{1}{2m} G^{2,0}+
 \frac{1}{2}m\omega^2 G^{0,2}\,.
\end{equation}
Any term that is not the standard kinetic energy evaluated for the momentum
expectation value may be considered as a contribution to an effective
potential, that is
\begin{equation} \label{VEffHarm}
 V_{\rm eff}= \frac{1}{2}m\omega^2 \langle\hat{q}\rangle^2+ \frac{1}{2m}
 G^{2,0}+  \frac{1}{2}m\omega^2 G^{0,2}\,.
\end{equation}

Further conditions may then be imposed in order to restrict states to more
specific regimes, for instance those near the ground state. For the harmonic
oscillator, the two conditions that states be stationary ($\dot{G}^{a,b}=0$)
and saturate the uncertainty relation
\begin{equation}
 G^{0,2}G^{2,0}-(G^{1,1})^2\geq \frac{\hbar^2}{4}
\end{equation}
are sufficient to restrict all the moments in $V_{\rm eff}$, with the result
that the quantum contribution is simply the zero-point energy
$\frac{1}{2}\hbar\omega$. For anharmonic oscillators, there are additional
terms coupling basic expectation values to the moments. They can be computed
in a semiclassical expansion and provide non-trivial,
$\langle\hat{q}\rangle$-dependent contributions to the effective potential. In
general, one can no longer use saturation of the uncertainty relation as one
of the conditions, but alternative and usually more complicated equations for
an interacting ground state are available.

Some of these results are reviewed and extended in Section~\ref{s:QM},
preceded by a formal discussion of an application of these methods in
quantum-field theory. In this article, we only sketch the extension of
canonical effective theory to quantum fields and focus on the
practical methods of deriving effective potentials in this way. A
comparison with the Coleman--Weinberg potential shows that we are able
to produce reliable results, which we subsequently apply to systems
with several degrees of freedom. (Details of canonical effective field
theory will be presented elsewhere.) In addition to a formal extension
to quantum field theory, further new ingredients in this paper include
canonical effective methods for fermionic variables and the associated
uncertainty relations and algebraic features.

\section{Quantum theory of a single scalar field}

When the methods and results leading to (\ref{VEffHarm}) are to be extended to
quantum field theory, an immediate problem is that moments and the effective
Hamiltonian are no longer well-defined, owing to the fact that they refer to
products of quantum fields evaluated at the same point. It turns out that it
is still possible to obtain meaningful results with a naive generalization of
(\ref{Moments}) and (\ref{EffHam}) to quantum fields. In this article we
focus on the practical aim of deriving effective potentials and discuss the
underlying conceptual questions elsewhere. We will therefore ignore
difficulties associated with products of quantum fields taken at the same
point. (We will, however, see examples of regularization in this framework.)

\subsection{Coleman--Weinberg potential from a canonical moment expansion} 

The classical Lagrangian for a massless scalar field (which we assume in
keeping with the original work of Coleman and Weinberg \cite{ColemanWeinberg})
with a quartic self-interaction is
\begin{equation}
{\rm L}=\int{\rm d}^3x\left[-\frac{1}{2}\eta^{\mu\nu}
\partial_{\mu}\phi(x)\partial_{\nu}\phi(x)-\lambda \phi^4(x)\right]\,,
\end{equation}
and the Hamiltonian
\begin{equation} \label{HamQFT}
{\rm H}=\int
{\rm d}^3x\left[\frac{\pi^2(x)}{2}+
\frac{1}{2}(\nabla\phi(x))^2+\lambda 
  \phi^4(x)\right]\,. 
\end{equation}
(We denote the momentum of $\phi(x)$ by $\pi(x)$, so that
$\{\phi(x),\pi(y)\}=\delta^3(x,y)$.)

A straightforward extension of (\ref{Moments}) to quantum fields leads us to
an `in-in' formalism, in which we calculate equal-time correlation
functions in some state. We introduce
\begin{eqnarray} \label{MomentsQFT}
G^{a,b}(x_1,\ldots,x_a;y_1,\ldots,y_b,t)&:=&
\bigg\langle\big(\hat{\pi}(x_1,t)-\langle\hat{\pi}(x_1,t)\rangle\big)\cdots
\big(\hat{\pi}(x_a,t)-\langle\hat{\pi}(x_a,t)\rangle\big)\times\nonumber\\ 
& &\quad\big(\hat{\phi}(y_1,t)-\langle\hat{\phi}(y_1,t)\rangle\big)\cdots
\big(\hat{\phi}(y_b,t)-\langle\hat{\phi}(y_b,t)\rangle\big)\bigg 
\rangle_{\rm Weyl} 
\end{eqnarray}
As before, the subscript `Weyl' stands for totally symmetric ordering. (In the
`in-in'-formalism, we are interested in equal-time correlation functions, so
that Weyl-ordering does not conflict with any time ordering.) We expect the
field-theory moments to have divergences whenever some of the $x_i$ or $y_j$
are identical. It is not obvious that such infinities can be ignored because
the analog of (\ref{EffHam}) applied to (\ref{HamQFT}) would include, for
instance, the moment $G^{0,2}(x,x)$. Nevertheless, we will see that the
variables (\ref{MomentsQFT}) are useful.  In the Hamiltonian, we also need
derivatives acting on different arguments of the moments, which are defined as
\begin{eqnarray} \label{MomentsDeriv}
&
&\nabla_{x_i}\nabla_{y_j}G^{a,b}(x_1,\ldots,x_a;y_1,\ldots,y_b,t):=
\nonumber\\ 
&
&\bigg\langle\big(\hat{\pi}(x_1,t)-\langle\hat{\pi}(x_1,t)\rangle\big)
\cdots\nabla_{x_i} 
\big(\hat{\pi}(x_i,t)-\langle\hat{\pi}(x_i,t)\rangle\big)\cdots\nonumber\\
&
&\times\big(\hat{\phi}(y_1,t)-\langle\hat{\phi}(y_1,t)\rangle\big)\cdots\nabla_{y_j} 
\big(\hat{\phi}(y_j,t)-\langle\hat{\phi}(y_j,t)\rangle\big)\cdots\bigg\rangle_{\rm 
  Weyl} \,.
\end{eqnarray}

Instead of working with the standard vacuum state of quantum field theory, or
any other specific Fock state, we now have an arbitrary state (indicated in
our notation just by expectation-value brackets) about which we calculate the
different correlation functions. Such a state may not only be non-Gaussian but
also mixed.  Thus, the calculations to follow are not limited to Fock states
as is the case for standard quantum field theory.

Following canonical effective methods in quantum mechanics, we write the
effective Hamiltonian as
\begin{eqnarray}
   {\rm H}_Q:=\langle \hat{\rm H}\rangle &=& \bigg\langle
   {\rm H}\big[\langle\hat{\phi}(x)\rangle+(\hat{\phi}(x)-
 \langle\hat{\phi}(x)\rangle),
\langle\hat{\pi}(x)\rangle+(\hat{\pi}(x)-
\langle\hat{\pi}(x)\rangle)\big]\bigg\rangle 
\end{eqnarray}
and perform an expansion by $\hat{\phi}(x)-\langle\hat{\phi}(x)\rangle$ and
$\hat{\pi}(x)-\langle\hat{\pi}(x)\rangle$. We have a finite series if ${\rm
  H}$ is polynomial, and a formal power-series expansion otherwise.
Abbreviating $\langle\hat{\pi}(x)\rangle\equiv\pi(x)$ and
$\langle\hat{\phi}(x)\rangle\equiv\phi(x)$, we obtain the effective
Hamiltonian as
\begin{eqnarray} \label{HQ4}
{\rm H}_Q&=&\frac{1}{2}\int {\rm d}^3x\bigg[\pi^2(x)+
G^{2,0}(x,x)+\nabla_x^2G^{0,2}(x,x)+\big(\nabla\phi(x)\big)^2\nonumber\\
& &+2\lambda\big\{\phi^4(x)+6\phi^2(x)G^{0,2}(x,x)+
4\phi(x)G^{0,3}(x,x,x)+G^{0,4}(x,x,x,x)\big\} \bigg]\,.
\end{eqnarray}
In general, we assume the Hamiltonian operator to be Weyl-ordered just as the
moments, so that there are no explicit $\hbar$-terms from re-ordering. In the
present case, of course, there are no ordering choices.

We may think of equal-$x$ moments as being regularized by point-splitting, so
that they are replaced by $G^{a,b}(x,y)$ multiplied with a smearing function
sharply peaked around $x=y$, in which we integrate over $x$ and $y$ to obtain
the effective Hamiltonian. Later on, we will make use of a precise subtraction
to remove divergences that would appear when the regulator is removed, that is
for $y\to x$ or when smearing functions approach a delta function. To
simplify our notation, we will not spell out the explicit point-splitting
regularizations. 

In order to solve the one-loop contribution to the effective potential (or
first order in $\hbar$), we do not need to solve for the $G^{0,3}$ and
$G^{0,4}$ terms: We expand around the free vacuum and therefore deal with
near-Gaussian states, the moments of which have a hierarchy in $\hbar$ that
goes as $G^{a,b}\propto\hbar^{(a+b)/2}$.  Thus, for contributions to first
order in $\hbar$, it is sufficient to restrict attention to the second-order
moments with $a+b=2$, that is $G^{0,2}, G^{2,0}$ and $G^{1,1}$. 
Our effective Hamiltonian turns into
\begin{eqnarray} \label{HQ42}
{\rm H}_Q&=&\frac{1}{2}\int {\rm d}^3x\bigg[\pi^2(x)+
G^{2,0}(x,x)+\nabla_x^2G^{0,2}(x,x)+\big(\nabla\phi(x)\big)^2\nonumber\\
& &+2\lambda\big\{\phi^4(x)+6\phi^2(x)G^{0,2}(x,x)+ O(\hbar^{3/2})
\big\} \bigg]\,.
\end{eqnarray}

\subsection{Solving for the relevant moments}

In order to calculate the Coleman--Weinberg potential in this canonical
formalism, we need to solve for the relevant moments in terms of $\phi(x)$ and
$\pi(x)$ and then insert the results in the expression (\ref{HQ42}).  We shall
solve the tree-level second-order moments as we are interested here only in
the one-loop contribution to the effective potential. Any higher-order
contributions to the moments will hence be ignored, and the equations of
motion as generated by ${\rm H}_{\rm Q}$ shall be truncated to include only
the lowest-order terms in $\hbar$. 

Some of our conditions correspond to moments of a state near the stationary
vacuum, and therefore require equations of motion. (Moments will have
vanishing time derivatives to leading order.) We can derive evolution
equations using, for instance, properties of Heisenberg operators. Time
derivatives of the moments are then equal to expectation values of commutators
with the Hamiltonian operator. It turns out to be easier to manage a system
based on mathematical methods of classical mechanics, namely phase spaces and
Poisson brackets. These methods are independent of which picture one prefers
for time evolution of operators or states --- Schr\"odinger, Heisenberg,
Dirac, or others.

We can view ${\rm H}_{\rm Q}$ as a Hamiltonian function on a phase space with
coordinates given by the expectation values $\phi(x)$ and $\pi(x)$, together
with the moments (\ref{MomentsQFT}) as new quantum degrees of freedom. As in
the case of effective quantum mechanics \cite{EffAc,Karpacz}, a Poisson
bracket can be defined for these variables (and functions of them) by
referring to the commutator of operators:
\begin{equation} \label{Poisson}
 \{\langle\hat{A}\rangle,\langle\hat{B}\rangle\}=
 \frac{\langle[\hat{A},\hat{B}]\rangle}{i\hbar}\,.
\end{equation}
If this definition is accompanied by the Leibniz rule for products of
expectation values, as they appear in the moments, a Poisson bracket is indeed
obtained.

For field operators, further divergences are introduced in this
Poisson bracket for expectation values of fields taken at the same
point. For instance, we have
\begin{eqnarray} \label{G02G20}
 \{G^{0,2}(x_1,x_2),G^{2,0}(y_1,y_2)\} &=& \delta^3(x_1-y_1) G^{1,1}(x_2,y_2)+
 \delta^3(x_1-y_2)G^{1,1}(x_2,y_1)\nonumber\\
 &&+ \delta^3(x_2-y_1) G^{1,1}(x_1,y_2)+
 \delta^3(x_2-y_2) G^{1,1}(x_1,y_1)\,.
\end{eqnarray}
However, these divergences are harmless as long as we compute Poisson brackets
with spatially integrated quantities such as the (regularized) effective
Hamiltonian (\ref{HQ4}). More interestingly, in some equations we can
eliminate all delta functions even if we do not explicitly regularize
(\ref{HQ4}). This is the case for the equations of motion of the second-order
moments, which are given by
\begin{eqnarray}\label{moments2}
\dot{G}^{0,2}(y,z,t)&=&G^{1,1}(y,z,t)+G^{1,1}(z,y,t) \label{G02dot}\\
\dot{G}^{1,1}(y;z,t)&=&G^{2,0}(y,z,t)-
[12\lambda\phi_0^2-\nabla^2_y]G^{0,2}(y,z,t) \label{G11dot}\\ 
\dot{G}^{2,0}(y,z,t)&=&-(12\lambda\phi_0^2-\nabla^2_z)G^{1,1}(y;z,t)+
-(12\lambda\phi_0^2-\nabla^2_y)G^{1,1}(z;y,t) \label{G20dot}
\end{eqnarray}
according to Hamiltonian equations of motion generated by (\ref{HQ4}).
No delta functions appear in these equations because the computation
of a Poisson bracket of the form $\{G^{a,b},{\rm H}_{\rm Q}\}$ with
the second-order version (\ref{HQ42}), using (\ref{G02G20}), always
gives integrated terms in which one $x$ of $G^{2,0}(x,x)$, say,
appears in a delta function from (\ref{G02G20}), and the only other
$x$ in a moment. For instance,
\[
 \frac{1}{2} \{G^{0,2}(y,z),\int{\rm d}^3x G^{2,0}(x,x)\}= \int{\rm d}^3x
 \left(\delta^3(y-x)G^{1,1}(z,x)+ \delta^3(z-x)G^{1,1}(y,x)\right)\,.
\]
However, if higher-order terms of (\ref{HQ4}) are included, starting with
$G^{0,3}(x,x,x)$, this convenient property will no longer be
realized. Nevertheless, a detailed analysis (to be presented elsewhere) shows
that infinities from equal-$x$ moments do not affect the system of equations.

It is important to remember, here and in what follows, that
there is no particular symmetry between the two arguments of $G^{1,1}(y,z,t)$
unlike that of $G^{0,2}(y,z,t)$ or $G^{2,0}(y,z,t)$, which are symmetric in
the spatial coordinates.

In order to find solutions of these equations in terms of $\phi(x)$ and
$\pi(x)$, we employ an adiabatic approximation: to the leading order, the
left-hand sides of the above equations are taken to be zero, rendering the
equations algebraic. (For our purposes, the leading-order approximation turns
out to be sufficient. Higher adiabatic orders do not contribute in the present
context because the vacuum expectation value $\langle\hat{\phi}\rangle$ will
be assumed to be independent of time and spatial coordinates.)

We first solve for $G^{1,1}$, which has to satisfy two conditions for the
right-hand sides of (\ref{G02dot}) and (\ref{G20dot}) to vanish:
\begin{eqnarray}
G^{1,1}(y,z)&=&-G^{1,1}(z,y)\\
(\nabla_y^2-\nabla_z^2)G^{1,1}(y,z)&=&0\,.
\end{eqnarray} 
If $G^{1,1}$ is to have a Fourier transform (or be absolutely integrable), the
only admissible solution is
\begin{equation}\label{G11}
 G^{1,1}(x,y)=0\,.
\end{equation} 
(Alternatively, one can use separation of variables to show that no other
solution satisfies both conditions.)

Only (\ref{G11dot}) remains to be solved, relating $G^{2,0}$ and $G^{0,2}$. By
symmetry, we can first formulate a condition just for $G^{0,2}$ because
\begin{equation}
G^{0,2}(y,z)=G^{0,2}(z,y)
\end{equation}
implies
\begin{equation} \label{G02b}
(\nabla_y^2-\nabla_z^2)G^{0,2}(y,z)=0\,.
\end{equation}
The variable $G^{0,2}$ should have a Fourier transform:
\begin{eqnarray}
G^{0,2}(y,z)=\int {\rm d}^3\vec{k}_y {\rm d}^3\vec{k}_z
f(\vec{k}_y,\vec{k}_z)e^{i[\vec{k}_y\cdot\vec{y}+\vec{k}_z\cdot\vec{z}]}
\end{eqnarray}
with $f(\vec{k}_y,\vec{k}_z)= f(\vec{k}_z,\vec{k}_y)$.  If we require
rotational invariance for an expansion around an isotropic (vacuum) state, the
only solution is $f(\vec{k}_y,\vec{k}_z)=g(k_z)\delta^3(\vec{k}_y-\vec{k}_z)$
with a function of a single variable $k_z=|\vec{k}_z|$,
or
\begin{equation}
 G^{0,2}(y,z)=\int {\rm d}^3\vec{k}_y \,\,
g(k_y)e^{i\vec{k}_y\cdot[\vec{y}-\vec{z}]}\,.
\end{equation}
The same function $g(k)$ determines
\begin{equation}
 G^{2,0}(y,z)=\int {\rm d}^3\vec{k}_y \,\,[12\lambda\phi_0^2+k_y^2]\,\,
g(k_y)e^{i\vec{k}_y\cdot[\vec{y}-\vec{z}]}
\end{equation}
by (\ref{G11dot}).

In order to solve for this function, we introduce a new condition for
the state to be close to the vacuum: We require that the second-order
moments should saturate the uncertainty relation. This latter equation
again requires care in a field-theory context, but it can be used in
order to obtain the crucial restriction on moments. Uncertainty
relations always follow from the Cauchy--Schwarz inequality
\begin{equation}
 \langle\hat{A}^{\dagger}\hat{A}\rangle
 \langle\hat{B}^{\dagger}\hat{B}\rangle\geq |\langle
 \hat{A}^{\dagger}\hat{B}\rangle|^2\,,
\end{equation}
valid for all states. For $\hat{A}=\hat{q}-\langle\hat{q}\rangle$ and
$\hat{B}=\hat{p}-\langle\hat{p}\rangle$, the standard uncertainty relation of
quantum mechanics follows:
\[
 G^{0,2}G^{2,0}\geq (G^{1,1})^2+\frac{\hbar^2}{4}\geq \frac{\hbar^2}{4}
\]
in our notation.

We now perform a formal calculation which gives an uncertainty relation for
field-theory moments: We first choose
$\hat{A}=\hat{\phi}(x)-\langle\hat{\phi}(x)\rangle$ and
$\hat{B}=\hat{\pi}(y)-\langle\hat{\pi}(y)\rangle$ and obtain
\begin{equation} \label{GxxGyy}
 G^{0,2}(x,x)G^{2,0}(y,y)\geq \frac{\hbar^2}{4}\delta^3(x-y)^2\,.
\end{equation}
Both sides are badly divergent and depend on how one regularizes
equal-$x$ moments and the square of a delta function.  Heuristically,
for $x\not=y$, the right-hand side is finite while the left-hand side
is infinite, so that the inequality is not very informative. However,
we can build on it and derive a more general inequality if we set
$\hat{A}=\frac{1}{2}(\hat{\phi}(x_1)-\langle\hat{\phi}(x_1)\rangle+
\hat{\phi}(x_2)-\langle\hat{\phi}(x_2)\rangle)$ and
$\hat{B}=\frac{1}{2}(\hat{\pi}(y_1)-\langle\hat{\pi}(y_1)\rangle+
\hat{\pi}(y_2)-\langle\hat{\pi}(y_2)\rangle)$. We obtain
\begin{eqnarray}
 &&\left(G^{0,2}(x_1,x_1)+ G^{0,2}(x_2,x_2)+ 2G^{0,2}(x_1,x_2)\right)
 \left(G^{2,0}(y_1,y_1)+ G^{2,0}(y_2,y_2)+2G^{2,0}(y_1,y_2)\right)\nonumber\\
&\geq&
\frac{\hbar^2}{4}\left(\delta^3(x_1-y_1)+\delta^3(x_1-y_2)+\delta^3(x_2-y_1)+ 
  \delta^3(x_2-y_2)\right)^2\,.
\end{eqnarray}
Multiplying all these terms and using (\ref{GxxGyy}) as an equality
(so as to subtract the worst divergences), we derive
\begin{eqnarray}
G^{0,2}(x_1,x_2)G^{2,0}(y_1,y_2)\geq
\frac{\hbar^2}{8}\left(\delta^3(x_1-y_1)\delta^3(x_2-y_2)+
\delta^3(x_1-y_2)\delta^3(x_2-y_1)\right) 
\end{eqnarray}
This inequality cannot be saturated in general, because for all four positions
different the 2-point functions may be non-zero but the right-hand side is
zero. However, we can require that the divergences on both sides are the same
(amounting to saturation of the delta-function contributions).

We therefore require that the equality
\begin{eqnarray}
G^{0,2}(x_1,x_2)G^{2,0}(y_1,y_2)=
\frac{\hbar^2}{8}\left(\delta^3(x_1-y_1)\delta^3(x_2-y_2)+
\delta^3(x_1-y_2)\delta^3(x_2-y_1)\right) 
\end{eqnarray}
holds in the sense of equal divergences on both sides.  (We are using
(\ref{G11}) in order to rule out any potential contribution from covariances.)
We can extract the leading divergences by setting $x_1=x_2$ and integrating
over $x_1$. Both sides are then equal if and only if
\begin{equation}
 g(k)=\frac{1}{2 (2\pi)^3}\frac{\hbar}{\sqrt{k^2+12\lambda\phi_0^2}}\,.
\end{equation}
Thus we have the full solution for the second-order moments as 
\begin{eqnarray}
G^{0,2}(y,z)&=&\frac{\hbar}{(2\pi)^3}\int 
\frac{{\rm d}^3\vec{k}}{2\sqrt{k^2+12\lambda\phi_0^2}}
e^{i\vec{k}\cdot(\vec{y}-\vec{z})}\\
G^{2,0}(y,z)&=&\frac{\hbar}{(2\pi)^3}\int 
\frac{{\rm d}^3\vec{k}}{2}\,\sqrt{k^2+12\lambda\phi_0^2}\,\,\,
e^{i\vec{k}\cdot(\vec{y}-\vec{z})}\,.
\end{eqnarray}

\subsection{The one-loop contribution to the effective potential}

The effective potential consists of all the terms in $H_{\rm Q}$,
other than the kinetic energy. To zeroth order, we just have
$\lambda\phi^4$.  We assume that the vacuum expectation value
$\phi_0:= \langle\widehat{\phi}\rangle$ is a constant, independent of
time and spatial coordinates. The one-loop contribution to the
potential in (\ref{HQ4}) or (\ref{HQ42}) is then given by
\begin{eqnarray}
&&\frac{1}{2} \left(G^{2,0}(x,x)+\nabla_x^2 G^{0,2}(x,x)\right)+ 
6\lambda \phi_0^2G^{0,2}(x,x)\nonumber\\ 
&=&\frac{\hbar}{2(2\pi)^3}\int {\rm d}^3 k
\left[\frac{\sqrt{k^2+12\lambda\phi_0^2}}{2}+\frac{k^2}{2\sqrt{k^2+
12\lambda\phi_0^2}}+ \frac{6\lambda\phi_0^2}{\sqrt{k^2+12\lambda\phi_0^2}}
 \right]\nonumber\\
&=&\frac{\hbar}{2(2\pi)^3}\int {\rm d}^3k\sqrt{k^2+12\lambda\phi_0^2}\,.
\end{eqnarray}
We may add an (infinite) constant $-\frac{\hbar}{2(2\pi)^3}\int {\rm d}^3k
|\vec{k}|$ to this one-loop contribution of the effective potential, and bring
it to the form
\begin{eqnarray} \label{VEff}
V_{\rm eff}(\phi_0)=\lambda \phi_0^4+\frac{\hbar}{2(2\pi)^3}\int {\rm d}^3k
  \left(-|\vec{k}|+\sqrt{k^2+12\lambda\phi_0^2}\right)\,.
\end{eqnarray}
In doing so we ensure that the effective potential is zero in the free limit
$\lambda\to0$. With the infinite constant we are therefore subtracting the
zero-point energies. This step corresponds to switching to a normal-ordered
Hamiltonian. (With an explicit point-splitting regularization, there would be
a finite subtraction before the regulator is removed.)

The usual form of the Coleman--Weinberg potential for a quartic
self-interaction is \cite{ColemanWeinberg}
\begin{eqnarray}
V_{\rm eff}(\phi_0)=\lambda \phi_0^4+\frac{i\hbar}{2}\int \frac{{\rm
    d}^4k}{(2\pi)^4}\log\left(1+\frac{12\lambda\phi_0^2}{-(k^0)^2+k^2}\right) \,.
\end{eqnarray}
It is possible to perform the integration over $k^0$ analytically, choosing a
suitable contour. The result,
\begin{eqnarray}
V_{\rm eff}(\phi_0)=\lambda \phi_0^4+\frac{\hbar}{2(2\pi)^3}\int {\rm d}^3k
  \left(-|\vec{k}|+\sqrt{k^2+12\lambda\phi_0^2}\right) \,,
\end{eqnarray}
agrees with our subtracted potential.

\section{The effective potential in quantum mechanics}
\label{s:QM}

Having established that effective potentials can be derived by a moment
expansion, we now illustrate the usefulness of the new method in a simplified
setting of quantum mechanics. For a single degree of freedom, our treatment is
closely related to the one of anharmonic oscillators in
\cite{EffAc,HigherTime}. In the following sections we will extend these
results to systems with two interacting degrees of freedom and to coupled
fermions, using a specific form suggested by particle physics.

If we eliminate the inhomogeneous modes by setting $\vec{k}=0$, we multiply
the integrand in (\ref{VEff}) with the momentum-space density $(2\pi)^3
\delta^3(\vec{k})$.  What remains of the one-loop correction to the effective
potential is then $\frac{1}{2}\hbar \sqrt{12\lambda\phi^2}=\frac{1}{2}\hbar
\sqrt{V''(\phi)}$. This result agrees with the anharmonic-oscillator model
which we now review. (See \cite{Karpacz,ReviewEff} for more details.)  We
assume a Hamiltonian given by
\begin{eqnarray}
H(\hat{p},\hat{q})=\frac{1}{2m}\hat{p}^2+V(\hat{q})\,.
\label{E3.1} 
\end{eqnarray}
The corresponding effective quantum Hamiltonian is
\begin{eqnarray}
H_Q(p,q,G^{a,b}):=\frac{1}{2m}(p^2+G^{2,0})+V(q)+
\sum_{n=2}^{\infty}\frac{1}{n!}V^{n}(q)G^{0,n}\,,  \label{E3.2}
\end{eqnarray} 
obtained as before by a formal Taylor expansion in
$\hat{q}-\langle\hat{q}\rangle$ and setting $q=\langle\hat{q}\rangle$,
$p=\langle\hat{p}\rangle$.  If the potential $\hat{V}$ is a polynomial in
$\hat{q}$, the expansion ends after a finite number of terms and is exact. (It
merely rearranges the polynomial $V(\hat{q})$ in $\hat{q}$ as a polynomial in
$\hat{q}-q$ with coefficients depending on $q$.) 

If we are interested only in corrections to the effective potential of first
order in $\hbar$, it is sufficient to restrict the expansion to second-order
moments. The effective Hamiltonian then becomes
\begin{eqnarray}
H_Q=\frac{1}{2m}(p^2+G^{2,0})+V(q)+\frac{1}{2}V''(q)G^{0,2} +
\mathcal{O}(\hbar^2)\,. \label{E3.3}
\end{eqnarray}
The equations of motion generated by this effective Hamiltonian (restricted to
first order in $\hbar$) for the (second order) moment terms are
\begin{eqnarray}
\dot{G}^{0,2}&=&\frac{2}{m}G^{1,1}\label{E3.4.1}\\
\dot{G}^{1,1}&=&\frac{1}{m}G^{2,0}-V''(q)G^{0,2}\label{E3.4.2}\\
\dot{G}^{2,0}&=& -2 V''(q)G^{1,1}\label{E3.4.3}
\end{eqnarray}
using again a Poisson bracket following from (\ref{Poisson}).  These three
equations are the only relevant ones for calculating the one-loop correction
to the effective potential. Higher-order moments in these equations can be
ignored because those contribute to higher loop corrections.

In order to solve the coupled ordinary differential equations, we again invoke
the adiabatic approximation to leading order, as all higher-order corrections
vanish for an expectation value $\langle\hat{q}\rangle$ independent of time.
In this case, both equations (\ref{E3.4.1}) and (\ref{E3.4.3}) give
$G^{1,1}=0$. The same equations will be used again at a later stage to give a
further constraint equation from the next adiabatic order.  Alternatively, we
could solve for $G^{1,1}$ from Eq.~(\ref{E3.4.1}) and look upon
Eq.~(\ref{E3.4.3}) as an independent consistency condition, which turns out to
be satisfied in this case. Of the equations which are used in deriving the
constraints later on, only certain subsets are independent while the rest
serve as consistency conditions at the leading adiabatic order. This feature
will turn out to be generic, as we shall see in the following sections.
 
The other equation, (\ref{E3.4.2}), tells us that
\begin{eqnarray}
G^{2,0}=mV''(q)G^{0,2}\,. \label{E3.5}
\end{eqnarray} 
Thus all the required moments, to this order, have either been solved for or
have been expressed in terms of $G^{0,2}$. In order to determine this one
remaining moment, we look at the constraint obtained by adding
Eqs.~(\ref{E3.4.1}) and (\ref{E3.4.3}):
\begin{eqnarray}
mV''(q)\dot{G}^{0,2}+\dot{G}^{2,0}=0 \label{E3.6}\,.
\end{eqnarray} 
(This equation goes beyond zeroth adiabatic order because the left-hand sides
in (\ref{E3.4.1}) and (\ref{E3.4.3}) are not set equal to zero. However, we
use it only as a consistency condition for solutions at zeroth adiabatic
order, not for a derivation of next-order contributions. For the latter, see
\cite{HigherTime}.)  Inserting Eq.~(\ref{E3.5}) in Eq.~(\ref{E3.6}), we obtain
the differential equation
\begin{eqnarray}
2V''(q)\dot{G}^{0,2}+V'''(q)\dot{q}G^{0,2}=0\label{E3.7}
\end{eqnarray}
with solution
\begin{eqnarray}
G^{0,2}=\frac{C}{\sqrt{V''(q)}}\,. \label{E3.8}
\end{eqnarray}
The constant $C$ remains to be determined. By requiring that the harmonic
limit $V(q) \rightarrow \frac{1}{2}m\omega^2q^2$ produces the well-known
expression for the fluctuation $G^{0,2}$, we find
$C=\frac{1}{2}\hbar/\sqrt{m}$.

Furthermore, this choice for the numerical constant implies that
the moments saturate the uncertainty relation,
\begin{eqnarray}
G^{0,2}G^{2,0}-\left(G^{1,1}\right)^2=\frac{\hbar^2}{4}\,. \label{E3.9}
\end{eqnarray}
In fact, as we shall see in the next section, the adiabatic constraint
equations sometimes form a rather complicated coupled set of differential
equations, which are difficult to solve directly. In those cases, we can more
easily solve for the free moments by saturating the uncertainty relation, and
then verify that the results indeed satisfy the system of constraints.  (By
requiring saturation of the uncertainty relation, we are looking for moments
of dynamical coherent states. Such states may not exist exactly, and indeed at
higher adiabatic order, it is in general not possible to saturate uncertainty
relations.)

We are now in a position to write down the effective potential for this
system. We have explicitly shown the calculation for the one-loop correction,
but with existing results of \cite{HigherTime} (following a similar procedure
but including higher order moments), we can write the effective potential to
two-loop order as
\begin{eqnarray}
V_{\rm eff}(q)=U(q)+\frac{\hbar}{2}\sqrt{\frac{V''(q)}{m}}+
\frac{\hbar^2}{8mV''(q)}\left[\frac{V''''(q)}{4}+
\frac{V'''(q)^2}{9V''(q)}\right]+\mathcal{O}(\hbar^3)\,.\label{E3.10}  
\end{eqnarray}

\section{Two coupled scalars}

We now consider a Hamiltonian in which two scalars are coupled to each other,
with some interaction terms that we will specify for resemblence with the
conformal standard model of \cite{ConfStand,ConfRenorm,ConfDiv}.  The
Hamiltonian operator for this system is given by
\begin{eqnarray}
H(\hat{p}_1,\hat{q}_1,\hat{p}_2,\hat{q}_2)=
\frac{1}{2m_1}\hat{p}_1^2+\frac{1}{2m_2}\hat{p}_2^2+
V(\hat{q}_1,\hat{q}_2)\,. \label{E4.1}
\end{eqnarray}  
The basic commutation relations are defined as 
\begin{eqnarray}
[\hat{q}_i,\hat{p}_j]=i\hbar\delta_{ij} \,\,\,\,\,({\rm with}\,\, i,j=1,2)
\end{eqnarray}

The moments are now of the form 
\begin{equation}
G^{a,b;c,d}:=
\bigg\langle\big(\hat{p}_1-\langle\hat{p}_1\rangle\big)^a
\big(\hat{q}_1-\langle\hat{q}_1\rangle\big)^b
\big(\hat{p}_2-\langle\hat{p}_2\rangle\big)^c
\big(\hat{q}_2-\langle\hat{q}_2\rangle\big)^d\bigg 
\rangle_{\rm Weyl} \,.
\end{equation}

\subsection{Moments}

As before, we want to calculate the effective potential to one-loop order, and
therefore we will only keep terms up to second order in
moments. The effective Hamiltonian, to this order, is
\begin{eqnarray}
H_Q(p_1,q_1,p_2,q_2,G^{a,b;c,d})&=&\frac{1}{2m_1}
\left(p_1^2+G^{2,0;0,0}\right)+
\frac{1}{2m_2}\left(p_2^2+G^{0,0;2,0}\right)\nonumber\\
&+&  V(q_1,q_2) +\frac{\partial^2V}{\partial q_1\partial q_2}(q_1,q_2)
G^{0,1;0,1}\nonumber\\
&+& \frac{1}{2}\frac{\partial^2V}{\partial q_1^2}(q_1,q_2)G^{0,2;0,0} + 
\frac{1}{2}\frac{\partial^2V}{\partial q_2^2}(q_1,q_2)G^{0,0;0,2}\,. 
\label{E4.2}
\end{eqnarray} 
From now on, we shall suppress the arguments of $ V(q_1, q_2):= V $. The
Hamiltonian generates equations of motion for the relevant second-order
moments:
\begin{eqnarray}
\dot{G}^{0,2;0,0}&=&\frac{2}{m_1}G^{1,1;0,0}\label{E4.3.1}\\
\dot{G}^{1,1;0,0}&=&\frac{1}{m_1}G^{2,0;0,0}-
\frac{\partial^2V}{\partial q_1^2}G^{0,2;0,0}-
\frac{\partial^2V}{\partial q_1\partial q_2}
G^{0,1;0,1}\label{E4.3.2}\\
\dot{G}^{2,0;0,0}&=&-2\frac{\partial^2V}{\partial q_1^2}
G^{1,1;0,0}-2\frac{\partial^2V}{\partial q_1\partial q_2}
G^{1,0;0,1}\label{E4.3.3}\\
\dot{G}^{0,0;0,2}&=&\frac{2}{m_2}G^{0,0;1,1}\label{E4.3.4}\\
\dot{G}^{0,0;1,1}&=&\frac{1}{m_2}G^{0,0;2,0}-
\frac{\partial^2V}{\partial q_2^2}G^{0,0;0,2}-
\frac{\partial^2V}{\partial q_1\partial q_2}
G^{0,1;0,1}\label{E4.3.5}\\
\dot{G}^{0,0;2,0}&=&-2\frac{\partial^2V}{\partial q_2^2}
G^{0,0;1,1}-2\frac{\partial^2V}{\partial q_1\partial q_2}
G^{0,1;1,0}\label{E4.3.6}\\
\dot{G}^{0,1;0,1}&=&\frac{1}{m_1}G^{1,0;0,1}
+\frac{1}{m_2}G^{0,1;1,0}\label{E4.3.7}\\
\dot{G}^{1,0;0,1}&=&\frac{1}{m_2}G^{1,0;1,0}-
\frac{\partial^2V}{\partial q_1^2}G^{0,1;0,1}-
\frac{\partial^2V}{\partial q_1\partial q_2} 
G^{0,0;0,2}\label{E4.3.8}\\
\dot{G}^{0,1;1,0}&=&\frac{1}{m_1}G^{1,0;1,0}-
\frac{\partial^2V}{\partial q_2^2}G^{0,1;0,1}-
\frac{\partial^2V}{\partial q_1\partial q_2}
G^{0,2;0,0}\label{E4.3.9}\\
\dot{G}^{1,0;1,0}&=&-\frac{\partial^2V}{\partial q_1^2}
G^{0,1;1,0}-\frac{\partial^2V}{\partial
    q_2^2}G^{1,0;0,1}-
\frac{\partial^2V}{\partial q_1\partial q_2}
\left(G^{0,0;1,1}-G^{1,1;0,0}\right)\label{E4.3.10}
\end{eqnarray}

Again we solve the equations to leading adiabatic order. For the effective
potential in (\ref{E4.2}), we need only the five moments $G^{2,0;0,0}$,
$G^{0,2;0,0}$, $G^{0,0;2,0}$, $G^{0,0;0,2}$ and $G^{0,1;0,1}$, but they cannot
be obtained directly from the set of equations without considering other
moments. We also note that there may be singular points in the resulting
potential. (Consistent adiabatic solutions may not exist for all field
values.) We can see the danger in the above system if we note that
Eqs.~(\ref{E4.3.2}) and (\ref{E4.3.5}) can be used to solve for $G^{2,0;0,0}$
and $G^{0,0;2,0}$ in terms of $G^{0,2;0,0}$ and $G^{0,0;0,2}$, respectively,
provided we know $G^{0,1;0,1}$. However, the latter moment can be determined
from the remaining equations only if an additional condition is satisfied:
(\ref{E4.3.8}) and (\ref{E4.3.9}) imply
\begin{equation}
 \left(m_2\frac{\partial^2V}{\partial q_1^2}- m_1 \frac{\partial^2V}{\partial
     q_2^2}\right) G^{0,1;0,1} = \frac{\partial^2V}{\partial q_1\partial q_2}
 \left(m_1G^{0,2;0,0}-m_2G^{0,0;0,2}\right)\,.
\end{equation}
If $m_2\partial^2V/\partial q_1^2- m_1 \partial^2V/\partial q_2^2=0$, the
moment remains undetermined.

It turns out to be easy to solve for some of the other covariance parameters.
From Eqs.~(\ref{E4.3.1}), (\ref{E4.3.3}), (\ref{E4.3.4}) and (\ref{E4.3.6}),
it immediately follows that
\begin{eqnarray}
G^{1,1;0,0}=0=G^{0,0;1,1}=G^{1,0;0,1}=G^{0,1;1,0}\,. \label{E4.4}
\end{eqnarray}
The equations used above to solve for these moments along with
Eqs.~(\ref{E4.3.7}) and (\ref{E4.3.10}), will give us further constraints
later on that will help us to fix two of the moments, which remain
undetermined after solving this homogeneous system of equations.

In more detail, we have a system of ten homogeneous equations in the leading
order adiabatic approximation. We manage to solve for eight of the ten moments
that show up in terms of two undetermined moments. This is a generic feature
of this method, just as in the single scalar case of Section~\ref{s:QM} where
we had a system of three homogeneous equations, using which we can solve for
only two of the three moments in terms of one undetermined one,
$G^{0,2}$. Analogously, in the present case, we will find that $G^{0,2;0,0}$
and $G^{0,0;0,2}$ remain undetermined upon solving the given system of
equations from only zeroth adiabatic order.

Using Eqs.~(\ref{E4.3.2}), (\ref{E4.3.5}), (\ref{E4.3.8}) and (\ref{E4.3.9}),
we can solve for the remaining moments in terms of $G^{0,2;0,0}$ and
$G^{0,0;0,2}$.
\begin{eqnarray}
G^{0,1;0,1}&=&\frac{\frac{\partial^2V}{\partial q_1\partial q_2}}
{m_2\frac{\partial^2V}{\partial q_1^2}-m_1\frac{\partial^2V}{\partial q_2^2}}
\left(m_1G^{0,2;0,0}-m_2G^{0,0;0,2}\right)\label{E4.5.1}\\
G^{2,0;0,0}&=&m_1\left[\left(\frac{\partial^2V}{\partial
      q_1^2}+\frac{\left(\frac{\partial^2V}{\partial q_1\partial
          q_2}\right)^2}{m_2\frac{\partial^2V}{\partial q_1^2}
-m_1\frac{\partial^2V}{\partial
    q_2^2}}\right)G^{0,2;0,0}\right.\nonumber\\
& &\,\,\,\,\,\,\,\,\,\,\,\,\,\,\,\,\,\,
\left.-\frac{\left(\frac{\partial^2V}{\partial q_1\partial q_2}\right)^2}
{m_2\frac{\partial^2V}{\partial q_1^2}- m_1\frac{\partial^2V}{\partial q_2^2}}
G^{0,0;0,2}\right]\label{E4.5.2}\\
G^{0,0;2,0}&=&m_2\left[\left(\frac{\partial^2V}{\partial
      q_2^2}+\frac{\left(\frac{\partial^2V}{\partial q_1\partial
          q_2}\right)^2}{m_1\frac{\partial^2V}{\partial q_2^2}
-m_2\frac{\partial^2V}{\partial q_1^2}}\right)G^{0,0;0,2}\right.
\nonumber\\
& &\,\,\,\,\,\,\,\,\,\,\,\,\,\,\,\,\,\,
\left.-\frac{\left(\frac{\partial^2V}{\partial q_1\partial q_2}\right)^2}
{m_1\frac{\partial^2V}{\partial
      q_1^2}-m_2\frac{\partial^2V}{\partial
      q_1^2}}G^{0,2;0,0}\right]\label{E4.5.3}\\ 
G^{1,0;1,0}&=&\frac{m_1m_2\frac{\partial^2V}{\partial q_1\partial
      q_2}}{m_2\frac{\partial^2V}{\partial
      q_1^2}-m_1\frac{\partial^2V}{\partial q_2^2}}
\left(\frac{\partial^2V}{\partial
    q_1^2}G^{0,2;0,0}-\frac{\partial^2V}{\partial q_2^2} 
G^{0,0;0,2}\right)\,.\label{E4.5.4}
\end{eqnarray}

We now have to find $G^{0,2;0,0}$ and $G^{0,0;0,2}$ before we can write down
an analytic expression for the effective potential, at one-loop order. We need
to solve a system of coupled differential equations, which can be obtained
from certain constraints, which we derive using the six Eqs.~(\ref{E4.3.1}),
(\ref{E4.3.3}), (\ref{E4.3.4}), (\ref{E4.3.6}), (\ref{E4.3.7}) and
(\ref{E4.3.10}), in analogy with (\ref{E3.7}) for one degree of freedom.
However we need only five of these equations at a time to derive a constraint
equation. Thus there are two independent euations that we can derive from the
original six. As before, we once again find that of the equations which are
used to derive the constraints, not all are independent at the leading
adiabatic order. We thus obtain consistency conditions, for instance
Eqs.~(\ref{E4.3.7}) and (\ref{E4.3.10}).

There are different choices available for the two independent constraints. We
may derive the two independent constraints as follows
\begin{eqnarray}
& &(\ref{E4.3.1})\times\frac{\partial^2V}{\partial q_1^2}+
(\ref{E4.3.3})\times\frac{1}{m_1}+
(\ref{E4.3.4})\times\frac{\partial^2V}{\partial q_2^2}\nonumber\\
& &+(\ref{E4.3.6})\times\frac{1}{m_2}+
(\ref{E4.3.7})\times\frac{\partial^2V}{\partial q_1\partial
    q_2}=0\label{E4.6.1}
\end{eqnarray}
and
\begin{eqnarray}
& &(\ref{E4.3.10})\times2\frac{\partial^2V}
{\partial q_1\partial q_2}-
(\ref{E4.3.6})\times\frac{\partial^2V}{\partial q_1^2}-
(\ref{E4.3.3})\times\frac{\partial^2V}{\partial q_2^2}\nonumber\\
& &+(\ref{E4.3.1})\times m_1\left[\left(\frac{\partial^2V}{\partial
      q_1\partial q_2}\right)^2-
\frac{\partial^2V}{\partial q_1^2}
\frac{\partial^2V}{\partial q_2^2}\right]\nonumber\\
& &+(\ref{E4.3.4})\times m_2\left[\left(\frac{\partial^2V}{\partial
      q_1\partial q_2}\right)^2-
\frac{\partial^2V}{\partial q_1^2}\frac{\partial^2V}{\partial q_2^2}
\right]=0\,.\label{E4.6.2}
\end{eqnarray}
These two equations translate into two conditions for the moments (and their
time derivatives). Once we insert solutions of all the other moments in terms
of $G^{0,2;0,0}$ and $G^{0,0;0,2}$, we end up with two coupled differential
equations for these two moments,
\begin{eqnarray}
& &2\left(\frac{\partial^2V}{\partial q_1^2}-\frac{\partial^2V}{\partial
    q_2^2}\right) \left[\frac{\partial^2V}{\partial q_1^2}
\left(\frac{\partial^2V}{\partial q_1^2}- \frac{\partial^2V}{\partial
    q_2^2}\right) +2\left(\frac{\partial^2V}{\partial q_1\partial
    q_2}\right)^2\right]\frac{{\rm d}X}{{\rm d}t}\nonumber\\
&+&2\left(\frac{\partial^2V}{\partial q_2^2}- \frac{\partial^2V}{\partial
    q_1^2}\right)\left[\frac{\partial^2V}{\partial q_2^2} 
\left(\frac{\partial^2V}{\partial q_2^2}- \frac{\partial^2V}{\partial
    q_1^2}\right)+ 2\left(\frac{\partial^2V}{\partial q_1\partial
    q_2}\right)^2\right]\frac{{\rm d}Z}{{\rm d}t}\nonumber\\ 
&+&\Bigg\{\left(\frac{\partial^2V}{\partial q_1^2}-
  \frac{\partial^2V}{\partial q_2^2}\right)^2\frac{{\rm d}}{{\rm d}t}
\frac{\partial^2V}{\partial q_1^2}
-4\left(\frac{\partial^2V}{\partial q_1\partial q_2}\right)^2
\left(\frac{{\rm d}}{{\rm d}t} \frac{\partial^2V}{\partial q_1^2}- \frac{{\rm
      d}}{{\rm d}t} \frac{\partial^2V}{\partial q_2^2}\right) \nonumber\\
& &+6\frac{\partial^2V}{\partial q_1\partial q_2}
\left(\frac{\partial^2V}{\partial q_1^2}- \frac{\partial^2V}{\partial
    q_2^2}\right)\frac{{\rm d}}{{\rm
      d}t} \frac{\partial^2V}{\partial q_1\partial q_2} \Bigg\}X\nonumber\\
&+&\Bigg\{\left(\frac{\partial^2V}{\partial q_2^2}-
  \frac{\partial^2V}{\partial q_1^2}\right)^2 \frac{{\rm d}}{{\rm d}t}
\frac{\partial^2V}{\partial q_2^2} 
-4\left(\frac{\partial^2V}{\partial q_1\partial q_2}\right)^2\left(\frac{{\rm
      d}}{{\rm d}t} \frac{\partial^2V}{\partial q_2^2}- \frac{{\rm d}}{{\rm
      d}t} \frac{\partial^2V}{\partial q_1^2}\right) \nonumber\\
& &+6\frac{\partial^2V}{\partial q_1\partial q_2} 
\left(\frac{\partial^2V}{\partial q_2^2}- \frac{\partial^2V}{\partial
    q_1^2}\right) 
\frac{{\rm d}}{{\rm d}t} \frac{\partial^2V}{\partial
  q_1\partial q_2} 
\Bigg\}Z=0
\end{eqnarray}
and
\begin{eqnarray}
& &2\left\{\left(\left(\frac{\partial^2V}{\partial q_1\partial q_2}\right)^2-
    \frac{\partial^2V}{\partial q_1^2} \frac{\partial^2V}{\partial q_2^2}\right)
\left(\frac{\partial^2V}{\partial q_1^2}- \frac{\partial^2V}{\partial
    q_2^2}\right) ^2\right\}\left(\frac{{\rm d} X}{{\rm d}t}+ \frac{{\rm d}
  Z}{{\rm d}t} \right) \nonumber\\
&-&\left(\frac{\partial^2V}{\partial q_1^2}- \frac{\partial^2V}{\partial
    q_2^2}\right) \Bigg\{-\left(\frac{\partial^2V}{\partial q_1\partial
    q_2}\right)^2 
\left(\frac{{\rm d}}{{\rm d}t} \frac{\partial^2V}{\partial q_1^2}+ 
\frac{{\rm d}}{{\rm d}t} \frac{\partial^2V}{\partial
  q_2^2}\right)\nonumber\\ 
& &+2\frac{\partial^2V}{\partial q_1\partial q_2}
\frac{\partial^2V}{\partial q_2^2}
 \frac{{\rm d}}{{\rm d}t} \frac{\partial^2V}{\partial q_1\partial q_2}
+\frac{\partial^2V}{\partial q_2^2}\left(\frac{\partial^2V}{\partial q_1^2}-
\frac{\partial^2V}{\partial q_2^2}\right)
\frac{{\rm d}}{{\rm d}t}  \frac{\partial^2V}{\partial q_1^2}
\Bigg\}X\nonumber\\
&-&\left(\frac{\partial^2V}{\partial q_2^2}- \frac{\partial^2V}{\partial
    q_1^2}\right) \Bigg\{-\left(\frac{\partial^2V}{\partial q_1\partial
    q_2}\right)^2 
\left(\frac{{\rm d}}{{\rm d}t} \frac{\partial^2V}{\partial q_1^2}+
\frac{{\rm d}}{{\rm d}t} \frac{\partial^2V}{\partial q_2^2}\right)\nonumber\\
& &+2\frac{\partial^2V}{\partial q_1\partial q_2}
\frac{\partial^2V}{\partial q_1^2}
\frac{{\rm d}}{{\rm d}t} \frac{\partial^2V}{\partial q_1\partial q_2}
+
\frac{\partial^2V}{\partial q_1^2}\left(\frac{\partial^2V}{\partial q_2^2}-
\frac{\partial^2V}{\partial q_1^2}\right)
\frac{{\rm d}}{{\rm d}t} \frac{\partial^2V}{\partial q_2^2}
\Bigg\}Z=0\nonumber
\end{eqnarray}
where we have abbreviated $X:=G^{0,2;0,0}$ and $Z:=G^{0,0;0,2}$. 

We are not aware of an easy way to solve these two coupled differential
equations for $G^{0,2;0,0}$ and $G^{0,0;0,2}$.  Instead, we use what we
learned from the example with one scalar, and try to solve for these two
moments by saturating the uncertainty relation. We will then insert the
solutions thus obtained in the two constraints to make sure that they are
legitimate solutions.

For two independent scalars, we have two uncertainty relations,
\begin{eqnarray}
G^{0,2;0,0}G^{2,0;0,0}-\left(G^{1,1;0,0}\right)^2&=&
\frac{\hbar^2}{4}\label{E4.7.1}\\
G^{0,0;0,2}G^{0,0;2,0}-\left(G^{0,0;1,1}\right)^2&=&
\frac{\hbar^2}{4}\,.\label{E4.7.2}
\end{eqnarray}
Using Eqs.~(\ref{E4.5.2}) and (\ref{E4.5.3}) (and (\ref{E4.4})), we can
substitute for $G^{2,0;0,0}$ and $G^{2,0;0,0}$ in terms of $G^{0,2;0,0}$ and
$G^{0,0;0,2}$ and thus solve for these two moments. For simplicity, let us
consider two particles of the same mass. (This assumption slightly simplifies
the expressions but is not necessary for an explicit solution.) Writing
$m_1=m_2=m$, we then have
\begin{eqnarray}
G^{0,2;0,0}&=&\frac{\hbar}{2\sqrt{m}}\times
\sqrt{\frac{1-\frac{\left(\frac{\partial^2V}{\partial
          q_1\partial q_2}\right)^2/\left(\frac{\partial^2V}{\partial
            q_2^2}
-\frac{\partial^2V}{\partial q_1^2}\right)}{\sqrt{\frac{\partial^2V}{\partial
  q_1^2} \frac{\partial^2V}{\partial q_2^2}
-\left(\frac{\partial^2V}{\partial q_1\partial q_2}\right)^2}}
}{\frac{\partial^2V}{\partial
      q_1^2}+\frac{\left(\frac{\partial^2V}{\partial q_1\partial
        q_2}\right)^2}{\frac{\partial^2V} 
{\partial q_1^2}- \frac{\partial^2V}{\partial q_2^2}}}}\label{E4.8.1}\\
G^{0,0;0,2}&=&\frac{\hbar}{2\sqrt{m}}\times
\sqrt{\frac{1-\frac{\left(\frac{\partial^2V}{\partial
          q_1\partial q_2}\right)^2/\left(\frac{\partial^2V}{\partial
            q_1^2}
-\frac{\partial^2V}{\partial q_2^2}\right)}{\sqrt{ \frac{\partial^2V}{\partial
  q_1^2} \frac{\partial^2V}{\partial q_2^2}
-\left(\frac{\partial^2V}{\partial q_1\partial q_2}\right)^2}}
}{\frac{\partial^2V}{\partial
      q_2^2}+\frac{\left(\frac{\partial^2V}{\partial q_1\partial
        q_2}\right)^2}{\frac{\partial^2V} 
{\partial q_2^2}-\frac{\partial^2V}{\partial q_1^2}}}} \,.
\label{E4.8.2}
\end{eqnarray}

Solving the quadratic equations for $G^{0,2;0,0}$ and $G^{0,0;0,2}$
given by the uncertainty relations, we have a freedom in choosing a
sign. As we will see in the particular example of the next section, we
need to choose the minus sign in front of the inner radical since the
other sign leads to a negative or imaginary value for $G^{2,0;0,0}$
and $G^{0,0;2,0}$, which is not permitted since these are
fluctuations. Having chosen the negative sign, we can express all the
moments in terms of the vacuum expectation values of the basic fields.

\subsection{Effective potential}

The moments inserted in the quantum Hamiltonian give the effective potential 
\begin{eqnarray}
V_{\rm eff}(q_1,q_2)&=& V(q_1,q_2)\\
&&+\frac{\hbar}{2\sqrt{m}}\left(\frac{2\left(\frac{\partial^2V}{\partial q_1\partial q_2}\right)^2}
{\frac{\partial^2V}{\partial q_1^2}-\frac{\partial^2V}{\partial q_2^2}}
+\frac{\partial^2V}{\partial q_1^2}\right)
\sqrt{\frac{1-\frac{\left(\frac{\partial^2V}{\partial q_1\partial q_2}\right)^2/\left(\frac{\partial^2V}{\partial q_1^2}
-\frac{\partial^2V}{\partial q_2^2}\right)}{\sqrt{\frac{\partial^2V}{\partial q_1^2}\frac{\partial^2V}{\partial q_2^2}
-\left(\frac{\partial^2V}{\partial q_1\partial q_2}\right)^2}}}
{\frac{\partial^2V}{\partial q_1^2}
+\frac{\left(\frac{\partial^2V}{\partial q_1\partial
      q_2}\right)^2}{\frac{\partial^2V}{\partial
    q_1^2}-\frac{\partial^2V}{\partial q_2^2}}}} \nonumber\\
&&+\frac{\hbar}{2\sqrt{m}}\left(\frac{2\left(\frac{\partial^2V}{\partial q_1\partial q_2}\right)^2}
{\frac{\partial^2V}{\partial q_2^2}-\frac{\partial^2V}{\partial q_1^2}}
+\frac{\partial^2V}{\partial q_2^2}\right)\sqrt{\frac{1-\frac{\left(\frac{\partial^2V}{\partial q_1\partial q_2}\right)^2/\left(\frac{\partial^2V}{\partial q_2^2}
-\frac{\partial^2V}{\partial q_1^2}\right)}{\sqrt{\frac{\partial^2V}{\partial q_1^2}\frac{\partial^2V}{\partial q_2^2}
-\left(\frac{\partial^2V}{\partial q_1\partial q_2}\right)^2}}}
{\frac{\partial^2V}{\partial q_2^2}
+\frac{\left(\frac{\partial^2V}{\partial q_1\partial
      q_2}\right)^2}{\frac{\partial^2V}{\partial q_2^2}- \frac{\partial^2V}{\partial q_1^2}}}}\,.\label{E4.9}
\end{eqnarray}
Although this expression looks complicated, it can be simplified slightly
once we insert a particular potential and evaluate the partial derivatives,
as illustrated now.

As already mentioned, the main interest for us in developing effective
potentials is to apply the methods to the conformal standard model of
\cite{ConfStand}. If we first turn off the fermion coupling to the
scalar, the resulting effective Hamiltonian is one for two scalar fields
with the same mass and a particular form of the potential:
\begin{eqnarray}
\hat{H}=\frac{1}{2m}\left(\hat{p}_1^2+\hat{p}_2^2\right)+
\lambda_1\hat{q}_1^4+\lambda_2\hat{q}_1^2\hat{q}_2^2+
\lambda_3\hat{q}_2^4 \,.\label{E5.1}
\end{eqnarray}  

Using the corresponding interaction terms in our expression for the effective
potential, we obtain
\begin{eqnarray}
V_{\rm eff}(q_1,q_2)&=& \lambda_1 q_1^4+\lambda_2q_1^2q_2^2+ \lambda_3 q_2^4 \nonumber\\
&&+\frac{\hbar}{2\sqrt{m}}\left(\frac{2g^2}{f_1-f_2}+f_1\right)\sqrt{\frac{1}{f_1+\frac{g^2}{f_1-f_2}}\left(1-\frac{g^2/\left(f_2-f_1\right)}{\sqrt{f_1f_2-g^2}}\right)}\nonumber\\
& & \,
+\frac{\hbar}{2\sqrt{m}}\left(\frac{2g^2}{f_2-f_1}+f_2\right)\sqrt{\frac{1}{f_2+\frac{g^2}{f_2-f_1}}\left(1-\frac{g^2/\left(f_1-f_2\right)}{\sqrt{f_1f_2-g^2}}\right)}\,.\label{E5.2} 
\end{eqnarray}
where we have defined the functions
$f_1(q_1,q_2):=12\lambda_1q_1^2+2\lambda_2q_2^2,\,\,\,
f_2(q_1,q_2):=12\lambda_3q_2^2+2\lambda_2q_1^2$ and
$g(q_1,q_2):=4\lambda_2q_1q_2$. 

\section{The conformal standard model}

In this final section we include the fermion terms in the Hamiltonian, in a
form motivated by \cite{ConfStand}:
\begin{eqnarray}
\hat{H}=\frac{1}{2m}\left(\hat{p}_1^2+\hat{p}_2^2\right)+\lambda_1\hat{q}_1^4+\lambda_2\hat{q}_1^2\hat{q}_2^2+\lambda_3\hat{q}_2^4+\frac{\alpha}{2}\hat{q}_1\left(\hat{\bar{\psi}}\hat{\psi}-\hat{\psi}\hat{\bar{\psi}} \,.
\right)\label{E6.1}
\end{eqnarray}
Such Hamiltonians require a further extension of the usual effective methods
of quantum mechanics, so that Grassmann variables describing fermions can be
included in the moments. 

\subsection{Moments of Grassmann variables}

We will use a straightforward extension of our notation for moments,
except that fermionic variables will be ordered totally
antisymmetrically. For better clarity, we write antisymmetric moments
with a tilde, such as $\tilde{G}^{0,0;0,0;1,1}$.  For the
corresponding effective equations, we need to extend the Poisson
brackets based on (\ref{Poisson}) to a graded version, making use of
anticommutators whenever two basic fermionic degrees of freedom
appear: for these variables, we have
$[\psi,\psi]_+=0=[\bar{\psi},\bar{\psi}]_+$ and
$[\bar{\psi},\psi]_+=i\hbar$.  Fermionic effective equations then
follow as before:
\begin{eqnarray}
\dot{\langle\hat{\mathcal{O}}\rangle}=
\{\langle\hat{\mathcal{O}}\rangle,H_{\rm Q}\}_{+} \,.
\end{eqnarray}

A classical model of a fermionic system can be obtained by using Grassmann-odd
variables $\psi$. Accordingly, the expectation values $\psi$ of fermion
operators in a canonical effective theory are Grassmann odd. Our present
scheme extends this well-known property to all moments which include an odd
number of fermion variables: Such moments, for instance $G^{1,0;0,0;1,0}$ are
odd Grassmann variables, while all moments with an even number of fermions are
Grassmann even.  The relative position between moment variables and fermionic
variables from expectation values is therefore important.

\subsection{Effective equations}

The corresponding effective Hamiltonian (up to second-order moments for the
one-loop contribution) is
\begin{eqnarray}
H_Q&=&\frac{1}{2m}\left(p_1^2+G^{2,0;0,0;0,0}+p_2^2+G^{0,0;2,0;0,0}\right)\nonumber\\
& & \,\,\,\, +\lambda_1\left(q_1^4+6q_1^2G^{0,2;0,0;0,0}\right)+\lambda_3\left(q_2^4+6q_2^2G^{0,0;0,2;0,0}\right)\nonumber\\
& & \,\,\,\, +\lambda_2\left(q_1^2q_2^2+q_1^2G^{0,0;0,2;0,0}+q_2^2G^{0,2;0,0;0,0}+4q_1q_2G^{0,1;0,1;0,0}\right)\nonumber\\
& & \,\,\,\, +\alpha \left(q_1\bar{\psi}\psi + q_1\tilde{G}^{0,0;0,0;1,1} + \bar{\psi}G^{0,1;0,0;0,1}+ G^{0,1;0,0;0,1}\psi\right)\,.\label{E6.2}
\end{eqnarray}
This Hamiltonian generates the following equations of motion for the moments
(up to first order in $\hbar$):
\begin{eqnarray}
\dot{G}^{0,2;0,0;0,0}&=&\frac{2}{m}G^{1,1;0,0;0,0}\label{E6.3.1}\\
\dot{G}^{1,1;0,0;0,0}&=&\frac{1}{m}G^{2,0;0,0;0,0}
-f_1 \,G^{0,2;0,0;0,0}-g \,G^{0,1;0,1;0,0}\nonumber\\
& &
-
\alpha\left(\bar{\psi} \,G^{0,1;0,0;0,1}+G^{0,1;0,0;1,0}\, \psi\right)
\label{E6.3.2}\\
\dot{G}^{2,0;0,0;0,0}&=&-2f_1 \,G^{1,1;0,0;0,0}-2g \,G^{1,0;0,1;0,0}\nonumber\\
& &
-2\alpha\left(\bar{\psi} \,G^{0,1;0,0;0,1}+G^{0,1;0,0;1,0}\, \psi\right)
\label{E6.3.3}\\
\dot{G}^{0,0;0,2;0,0}&=&\frac{2}{m}G^{0,0;1,1;0,0}\label{E6.3.4}\\
\dot{G}^{0,0;1,1;0,0}&=&\frac{1}{m}G^{0,0;2,0;0,0}
-f_2 \,G^{0,0;0,2;0,0}-g \,G^{0,1;0,1;0,0}\label{E6.3.5}\\
\dot{G}^{0,0;2,0;0,0}&=&-2f_2 \,G^{0,0;1,1;0,0}
-2g \,G^{0,1;1,0;0,0}\label{E6.3.6}\\
\dot{G}^{0,1;0,1;0,0}&=&\frac{1}{m}\left(G^{1,0;0,1;0,0}-G^{0,1;1,0;0,0}\right)\label{E6.3.7}\\
\dot{G}^{1,0;0,1;0,0}&=&\frac{1}{m}G^{1,0;1,0;0,0}-f_1\,G^{0,1;0,1;0,0}
-g \,G^{0,0;0,2}\nonumber\\
& &
-\alpha\left(\bar{\psi} \,G^{0,0;0,1;0,1}+G^{0,0;0,1;1,0}\, \psi\right) 
\label{E6.3.8}\\
\dot{G}^{0,1;1,0;0,0}&=&\frac{1}{m}G^{1,0;1,0;0,0}
-f_2 \,G^{0,1;0,1;0,0}-g \,G^{0,2;0,0;0,0}\label{E6.3.9}\\
\dot{G}^{1,0;1,0;0,0}&=&-f_1 \,G^{0,1;1,0;0,0}-f_2 \,G^{1,0;0,1;0,0}
-g \,\left(G^{0,0;1,1;0,0}+G^{1,1;0,0;0,0}\right)\nonumber\\ & &
-\alpha\left(\bar{\psi} \,G^{0,0;1,0;0,1}+G^{0,0;1,0;1,0}\, \psi\right) 
\label{E6.3.10}\\
\dot{G}^{0,1;0,0;1,0}&=&\frac{1}{m}G^{1,0;0,0;1,0}+
i\alpha\left(\bar{\psi} \,G^{0,2;0,0;0,0}+q_1 \,G^{0,1;0,0;1,0}\right)
\label{E6.3.11}\\
\dot{G}^{0,1;0,0;0,1}&=&\frac{1}{m}G^{1,0;0,0;0,1}-
i\alpha\left(G^{0,2;0,0;0,0} \,\psi+q_1 \,G^{0,1;0,0;0,1}\right)
\label{E6.3.12}\\
\dot{G}^{1,0;0,0;1,0}&=&-f_1 \,G^{0,1;0,0;1,0}-g \,G^{0,0;0,1;1,0}\nonumber\\
& &+
i\alpha\left(q_1\,G^{1,0;0,0;1,0}+\bar{\psi} \,G^{1,1;0,0;0,0}-i\bar{\psi} \,\tilde{G}^{0,0;0,0;1,1}\right)\label{E6.3.13}\\
\dot{G}^{1,0;0,0;1,0}&=&-f_1 \,G^{0,1;0,0;0,1}-g \,G^{0,0;0,1;0,1}\nonumber\\
& &
-i\alpha\left(q_1\,G^{1,0;0,0;0,1}+ G^{1,1;0,0;0,0} \,\psi + i\tilde{G}^{0,0;0,0;1,1}\,\psi\right)\label{E6.3.14}\\
\dot{G}^{0,0;0,1;1,0}&=&\frac{1}{m}G^{0,0;1,0;1,0}
+i\alpha\left(\bar{\psi} \,G^{0,1;0,1;0,0} +q_1 \,G^{0,0;0,1;1,0}\right)
\label{E6.3.15}\\
\dot{G}^{0,0;0,1;0,1}&=&\frac{1}{m}G^{0,0;1,0;0,1}
-i\alpha\left(G^{0,1;0,1;0,0} \,\psi+q_1 \,G^{0,0;0,1;0,1}\right)
\label{E6.3.16}\\
\dot{G}^{0,0;1,0;1,0}&=&-f_2 \,G^{0,0;0,1;1,0}-g \,G^{0,1;0,0;1,0}\nonumber\\
& &
+i\alpha\left(q_1\,G^{0,0;1,0;1,0}+\bar{\psi} \,G^{0,1;1,0;0,0}\right)
\label{E6.3.17}\\
\dot{G}^{0,0;1,0;0,1}&=&-f_2 \,G^{0,0;0,1;0,1}-g \,G^{0,1;0,0;0,1}\nonumber\\
& &
-i\alpha\left(q_1\,G^{0,0;1,0;0,1}+ G^{0,1;1,0;0,0} \,\psi\right)
\label{E6.3.18}\\
\dot{\tilde{G}}^{0,0;0,0;1,1}&=&i\alpha\left(\bar{\psi}
  \,G^{0,1;0,0;0,1}-G^{0,1;0,0;1,0}\,\psi\right) \,.\label{E6.3.19}
\end{eqnarray}

Once we apply the adiabatic approximation, we are left with a system of
nineteen homogeneous equations. From these equations, we are able to solve for
sixteen of the moments in terms of three undetermined moments
($G^{0,2;0,0;0,0}, G^{0,0;0,2;0,0}$ and $\tilde{G}^{0,0;0,0;1,1}$). These
latter will be fixed by requiring that they satisfy certain constraint
equations as before. Equivalently, we solve for them by saturating the three
uncertainty relations for the three canonical pairs, and then making sure that
the solutions satisfy all constraint equations.

Looking at Eqs. (\ref{E6.3.1}), (\ref{E6.3.4}), (\ref{E6.3.6}) and
(\ref{E6.3.7}), we can immediately set $G^{1,1;0,0;0,0} = G^{0,0;1,1;0,0} =
G^{0,1;1,0;0,0} = G^{1,0;0,1;0,0} = 0$. If we look at a subset of these
equations, Eqs. (\ref{E6.3.11})--(\ref{E6.3.19}), then we can solve for the
Grassmann-odd moments in terms of the rest of the moments:
\begin{eqnarray}
D G^{0,1;0,0;1,0}&=& -\alpha\bar{\psi}\Bigg[f_2\left(m\alpha q_1 G^{0,2;0,0;0,0}+\tilde{G}^{0,0;0,0;1,1}\right)\nonumber\\& &
\,\,\,\,\,\,\,\,\,-m\alpha q_1\left(gG^{0,1;0,1;0,0}+\alpha q_1 \left\{\tilde{G}^{0,0;0,0;1,1}+m\alpha q_1G^{0,2;0,0;0,0}\right\}\right)\Bigg]\label{E6.4.1}\\
D G^{0,1;0,0;0,1}&=& -\alpha\psi\Bigg[f_2\left(m\alpha q_1 G^{0,2;0,0;0,0}+\tilde{G}^{0,0;0,0;1,1}\right)\nonumber\\& &
\,\,\,\,\,\,\,\,\,-m\alpha q_1\left(gG^{0,1;0,1;0,0}+\alpha q_1 \left\{\tilde{G}^{0,0;0,0;1,1}+m\alpha q_1G^{0,2;0,0;0,0}\right\}\right)\Bigg]\label{E6.4.2}\\
D G^{1,0;0,0;1,0}&=&-im\alpha\bar{\psi}\Bigg[\alpha q_1\left(\left\{m\alpha^2q_1^2-f_2\right\}\tilde{G}^{0,0;0,0;1,1}+gm\alpha q_1G^{0,1;0,1;0,0}\right)\nonumber\\& &
\,\,\,\,\,\,\,\,\,\,\,\,\,\,\,\,\,\,+\left(g^2-f_1f_2+m\alpha^2 q_1^2f_1\right)G^{0,2;0,0;0,0}\Bigg]\label{E6.4.3}\\
D G^{1,0;0,0;0,1}&=&im\alpha\psi\Bigg[\alpha q_1\left(\left\{m\alpha^2q_1^2-f_2\right\}\tilde{G}^{0,0;0,0;1,1}+gm\alpha q_1G^{0,1;0,1;0,0}\right)\nonumber\\
& &\,\,\,\,\,\,\,\,\,\,\,\,\,\,\,\,\,+\left(g^2-f_1f_2+m\alpha^2 q_1^2f_1\right)G^{0,2;0,0;0,0}\Bigg]\label{E6.4.4}\\
D G^{0,0;0,1;1,0}&=&\alpha\bar{\psi}\left[g\tilde{G}^{0,0;0,0;1,1}+m\alpha q_1\left(\left\{m\alpha^2q_1^2-f_1\right\}G^{0,1;0,1;0,0}+gG^{0,2;0,0;0,0}\right)\right]\label{E6.4.5}\\
D G^{0,0;0,1;0,1}&=&\alpha\psi\left[g\tilde{G}^{0,0;0,0;1,1}+m\alpha q_1\left(\left\{m\alpha^2q_1^2-f_1\right\}G^{0,1;0,1;0,0}+gG^{0,2;0,0;0,0}\right)\right]\label{E6.4.6}\\
D G^{0,0;1,0;1,0}&=&-im\alpha\bar{\psi}\Bigg[\alpha g q_1\tilde{G}^{0,0;0,0;1,1}+\left(g^2-f_1f_2\right)G^{0,1;0,1;0,0}\nonumber\\
& &\,\,\,\,\,\,\,\,\,\,\,\,\,\,\,\,\,\,\,\,+m\alpha^2q_1^2\left(f_2G^{0,1;0,1;0,0}+g G^{0,2;0,0;0,0}\right)\Bigg]\label{E6.4.7}\\
D G^{0,0;1,0;0,1}&=&im\alpha\psi\Bigg[\alpha g
     q_1\tilde{G}^{0,0;0,0;1,1}+\left(g^2-f_1f_2\right)G^{0,1;0,1;0,0}\nonumber\\
& &\,\,\,\,\,\,\,\,\,\,\,\,\,\,\,\,+m\alpha^2q_1^2\left(f_2G^{0,1;0,1;0,0}+g
       G^{0,2;0,0;0,0}\right)\Bigg] \,.\label{E6.4.8}
\end{eqnarray}
where $D=g^2-\left(f_1-m\alpha^2q_1^2\right)\left(f_2-m\alpha^2q_1^2\right)$.

At this point, Eqs. (\ref{E6.3.3}), (\ref{E6.3.10}) and (\ref{E6.3.19}) can be
used as consistency checks for the above solutions. Once we use in the values
of the trivial moments in these equations, they reduce to the form
\begin{eqnarray}
\psi G^{0,0;1,0;1,0} +\bar{\psi}G^{0,0;1,0;0,1}&=&0\label{E6.5.1}\\
\psi G^{1,0;0,0;1,0} +\bar{\psi}G^{1,0;0,0;0,1}&=&0\label{E6.5.2}\\
\psi G^{0,1;0,0;1,0} -\bar{\psi}G^{0,1;0,0;0,1}&=&0 \,.\label{E6.5.3}
\end{eqnarray}
Inserting the solutions for the moments, it can be shown that they satisfy
these constraints identically. In order to evaluate the left-hand side of the
above set of equations, it is important to remember at this point that the
$\psi, \bar{\psi}$ and the Grassmann-odd moments are anticommuting objects.

The rest of the equations can then be used to solve for the Grassman-even
moments in terms of $G^{0,2;0,0;0,0}, G^{0,0;0,2;0,0}$ and
$\tilde{G}^{0,0;0,0;1,1}$.
\begin{eqnarray}
G^{2,0;0,0;0,0}&=&m\Bigg[f_1G^{0,2;0,0;0,0}-\frac{2\bar{\psi}\psi m\alpha^3q_1(f_2-m\alpha^2q_1^2)}{D}\nonumber\\
& &+\frac{g^2\left(1+\frac{2\bar{\psi}\psi m\alpha^3q_1}{D}\right)\left(G^{0,0;0,2;0,0}-G^{0,2;0,0;0,0}+\frac{2\bar{\psi}\psi\alpha^2\left(\tilde{G}^{0,0;0,0;1,1}+m\alpha q_1G^{0,2;0,0;0,0}\right)}{D}\right)}{f_2-f_1+\frac{2\bar{\psi}\psi m\alpha^3q_1\left(f_1-m\alpha^2q_1^2\right)}{D}}\nonumber\\
& &-\frac{2\bar{\psi}\psi\left(f_2-m\alpha^2q_1^2\right)\tilde{G}^{0,0;0,0;1,1}}{D}\Bigg]\label{E6.6.1}\\
G^{0,0;2,0;0,0}&=&m f_2G^{0,0;0,2;0,0}\nonumber\\
& &+\frac{mg^2\left(G^{0,0;0,2;0,0}-G^{0,2;0,0;0,0}+\frac{2\bar{\psi}\psi \alpha^2\left(\tilde{G}^{0,0;0,0;1,1}+m\alpha q_1G^{0,2;0,0;0,0}\right)}{D}\right)}{f_2-f_1+\frac{2\bar{\psi}\psi\alpha^3 mq_1\left(f_1-m\alpha^2q_1^2\right)}{D}}\label{E6.6.2}\\
G^{0,1;0,1;0,0}&=&\frac{g^2\left(G^{0,0;0,2;0,0}-G^{0,2;0,0;0,0}+\frac{2\bar{\psi}\psi \alpha^2\left(\tilde{G}^{0,0;0,0;1,1}+m\alpha q_1G^{0,2;0,0;0,0}\right)}{D}\right)}{f_2-f_1+\frac{2\bar{\psi}\psi\alpha^3 mq_1\left(f_1-m\alpha^2q_1^2\right)}{D}}\label{E6.6.3}\\
G^{1,0;1,0;0,0}&=&mg\Bigg[G^{0,2;0,0;0,0}\nonumber\\
& &+\frac{f_2\left(G^{0,0;0,2;0,0}-G^{0,2;0,0;0,0}+\frac{2\bar{\psi}\psi \alpha^2\left(\tilde{G}^{0,0;0,0;1,1}+m\alpha q_1G^{0,2;0,0;0,0}\right)}{D}\right)}{f_2-f_1+\frac{2\bar{\psi}\psi\alpha^3 mq_1\left(f_1-m\alpha^2q_1^2\right)}{D}}\Bigg]\label{E6.6.4}
\end{eqnarray} 
Finally, we should solve for these three moments $G^{0,2;0,0;0,0},
G^{0,0;0,2;0,0}$ and $\tilde{G}^{0,0;0,0;1,1}$ by saturating the three
uncertainty relations. Once this step is completed all the moments are
expressed in terms of the field expectation values. The solution for these
moments, obtained by solving the uncertainty relations, must satisfy
certain constraints as before. One such constraint equation is of the form
\begin{eqnarray}
 & &\frac{1}{m}\dot{G}^{2,0;0,0;0,0}+f_1\dot{G}^{0,2;0,0;0,0}+\frac{1}{m}\dot{G}^{0,0;2,0;0,0}+f_2\dot{G}^{0,0;0,2;0,0}+2g\dot{G}^{0,1;0,1;0,0}\nonumber\\
&
 & + 2\alpha\left[\dot{G}^{0,1;0,0;1,0}\psi+\bar{\psi}\dot{G}^{0,1;0,0;0,1}+
q_1\dot{\tilde{G}}{}^{0,0;0,0;1,1}\right]=0\,. \label{E6.7.1}
\end{eqnarray}
Since we have the solution of all the relevant moments in terms of the three
moments $G^{0,2;0,0;0,0}, G^{0,0;0,2;0,0}$ and $\tilde{G}^{0,0;0,0;1,1}$,
(\ref{E6.7.1}) gives us a constraint on these three moments. Similarly there
are two other independent constraints that we can derive for these moments.

\subsection{The Uncertainty Relations}

Two of the three uncertainty realtions that we saturate in order to get
the solutions for the three remaining moments can be written in terms of the
bosonic moments.
\begin{eqnarray}
 G^{0,2;0,0;0,0}G^{2,0;0,0;0,0}&=&\frac{\hbar^2}{4}\label{E6.8.1}\\
 G^{0,0;0,2;0,0}G^{0,0;2,0;0,0}&=&\frac{\hbar^2}{4}\,.\label{E6.8.2}
\end{eqnarray}
Since we have already solved for $G^{2,0;0,0;0,0}$ and $G^{0,0;2,0;0,0}$ in
terms of $G^{0,2;0,0;0,0}, G^{0,0;0,2;0,0}$ and $\tilde{G}^{0,0;0,0;1,1}$, we
get two simultaneous quadratic equations in the three undetermined
moments. However, the third uncertainty relation involving the fermionic field
can be saturated to solve for $\tilde{G}^{0,0;0,0;1,1}$ independent of all
other moments in terms of the fermionic fields:
\begin{eqnarray}
 \tilde{G}^{0,0;0,0;1,1}=2\bar{\psi}\psi \pm \frac{1}{2}\hbar\,.\label{E6.8.3}
\end{eqnarray}
(Strictly speaking, this equation is not an uncertainty relation for
fluctuations, but it follows in an analogous way.)

Next we can solve for $G^{0,2;0,0;0,0}$ and $G^{0,0;0,2;0,0}$ in terms of
$q_1, q_2$ using the above relations. Although we manage to solve for all the
moments in terms of the field expectation values, the explicit form of the
solution in this case looks extremely cumbersome and lengthy. Thus we refrain
from quoting the explicit solutions here. The form of the effective potential
up to $\mathcal{O}(\hbar)$, expressed in terms of these moments can then be
written as
\begin{eqnarray}\label{EffPotn}
 V_{\rm eff}&=&\lambda_1q_1^4+ \lambda_3q_2^4 +\lambda_2q_1^2q_2^2 + \alpha q_1\bar{\psi}\psi\nonumber\\
& &\,\,\,\,\,\,\,+\frac{\hbar}{2\sqrt{m}}\Bigg[G^{2,0;0,0;0,0}+G^{0,0;2,0;0,0} + f_1 G^{0,2;0,0;0,0}
+f_2 G^{0,0;0,2;0,0} +2 g G^{0,1;0,1;0,0}\nonumber\\
& &\,\,\,\,\,\,\,\,\,\,\,\,\,\,\,\,\,\,\,\,\,\,\,\,\,+\alpha\left(q_1\tilde{G}^{0,0;0,0;1,1} + \bar{\psi}G^{0,1;0,0;0,1}+ G^{0,1;0,0;0,1}\psi \right)\Bigg] \,.
\end{eqnarray}

Of course, it can be checked that setting $\alpha=0$ gives us back the
effective potential for the two-field model described above. Obviously to get
the exact expression of the effective potential in terms of the field
expectation values, we now need to expand the moments in the above equation in
terms of their exact solutions. Although we do not give explicit formulae for
the effective potential in this case, we have described the general algorithm
for calculating it in this canonical formalism. As before, we need to solve
quartic equations, but they are not simple and with Grassmann-odd coefficients
computer algebra methods cannot be applied straightforwardly.  Nevertheless,
in models in which solving for the explicit version of the one-loop effective
potential might become difficult, one can still use numerical methods to solve
for these equations.

\section{Conclusions}

We have presented several extensions of the canonical theory of effective
equations in preparation for an application to quantum-field theory. We have
considered suitable definitions for effective equations of $n$-point
functions, but no explicit regularization has been provided. Still, we
obtained several useful results which indicate that the usual divergences can
safely be cured in this setting, at least regarding the questions posed in
this paper. The results we obtained are fully consistent  with established
methods, including details of the Coleman--Weinberg potential. 

Some of the new general constructions provided here include uncertainty
relations for field theories and a definition of effective equations for
fermions. Several examples have shown that some results can be obtained
analytically, while the more complicated type of equations for which we have
not been able to find closed solutions seems to be amenable to numerical
methods. Some open computational problems remain before our scheme can be
turned into a fully automated code to produce numerical data on effective
potentials. At the present stage, however, such a program seems to be feasible
with the methods of this paper.

\section*{Acknowledgements}

This work was supported in part by NSF grant PHY-1307408.


\begin{thebibliography}{1}

\bibitem{EffAc}
M.\ Bojowald and A.\ Skirzewski,
\newblock Effective Equations of Motion for Quantum Systems,
\newblock {\em Rev.\ Math.\ Phys.} 18 (2006) 713--745, [math-ph/0511043]

\bibitem{Karpacz}
M.\ Bojowald and A.\ Skirzewski,
\newblock Quantum Gravity and Higher Curvature Actions,
\newblock {\em Int.\ J.\ Geom.\ Meth.\ Mod.\ Phys.} 4 (2007) 25--52,
  [hep-th/0606232],
\newblock Proceedings of ``Current Mathematical Topics in Gravitation and
  Cosmology'' (42nd Karpacz Winter School of Theoretical Physics), Ed.\
  Borowiec, A.\ and Francaviglia, M.

\bibitem{HigherTime}
M.\ Bojowald, S.\ Brahma, and E.\ Nelson,
\newblock Higher time derivatives in effective equations of canonical quantum
  systems,
\newblock {\em Phys.\ Rev.\ D} 86 (2012) 105004, [arXiv:1208.1242]

\bibitem{ColemanWeinberg}
S.\ Coleman and E.\ Weinberg,
\newblock Radiative corrections as the origin of spontaneous symmetry breaking,
\newblock {\em Phys.\ Rev.\ D} 7 (1973) 1888--1910

\bibitem{ReviewEff}
M.\ Bojowald,
\newblock Quantum Cosmology: Effective Theory,
\newblock {\em Class.\ Quantum Grav.} 29 (2012) 213001, [arXiv:1209.3403]

\bibitem{ConfStand}
K.~A.\ Meissner and H.\ Nicolai,
\newblock Conformal Symmetry and the Standard Model,
\newblock {\em Phys.\ Lett.\ B} 648 (2007) 312--317, [hep-th/0612165]

\bibitem{ConfRenorm}
K.~A.\ Meissner and H.\ Nicolai,
\newblock Renormalization group and effective potential in classically
  conformal theories,
\newblock {\em Acta Phys.\ Pol.} 40 (2009) 2737--2752, [arXiv:0809.1338]

\bibitem{ConfDiv}
K.~A.\ Meissner and H.\ Nicolai,
\newblock Effective action, conformal anomaly and the issue of quadratic
  divergences,
\newblock {\em Phys.\ Lett.\ B} 660 (2008) 260--266, [arXiv:0710.2840]

\end{thebibliography}

\end{document}